\documentclass[12pt]{article}

\usepackage[dvips]{epsfig}
\usepackage{amsfonts}
\usepackage{amscd}
\usepackage{amsmath}
\usepackage{pstricks}
 \font\tensym=msbm10
 \font\sevensym=msbm7
 \font\fivesym=msbm5

 \font\tengoth=eufb10
 \font\sevengoth=eufb7
 \font\fivegoth=eufb5

\newfam\symfam
\textfont\symfam=\tensym
\scriptfont\symfam=\sevensym
\scriptscriptfont\symfam=\fivesym

\newfam\gothfam
\textfont\gothfam=\tengoth
\scriptfont\gothfam=\sevengoth
\scriptscriptfont\gothfam=\fivegoth

\def\hs{\hbox to 3mm{}}
\def\hhs{\hbox to 5cm{}}

\def\JPicScale{0.8}\ifx\JPicScale\undefined\def\JPicScale{1}\fi

\def\C{\mathbb{C}}

\def\N{\mathbb{N}}
\def\Z{\mathbb{Z}}

\def\H{\mathcal{H}}

\def\HWS{{\cal H}_{WS}}
\def\1H{\mathbf{1}_{\HWS}}

\def\al{\alpha}
\def\be{\beta}

\def\2m#1#2{(#2 #1)}
\def\3m#1#2#3{(#3 #2 #1)}

\def\ncp#1#2{#1\langle #2\rangle}

\def\scal#1#2{\langle #1 | #2 \rangle}

\def\ra{\rightarrow}

\def\adots{\mathinner{\mkern2mu\raise1pt\hbox{.}
\mkern3mu\raise4pt\hbox{.}\mkern1mu\raise7pt\hbox{.}}}
\def\binomial#1#2{\begin{pmatrix}#1 \\ #2\end{pmatrix}}

\def\up#1{\raise 1ex\hbox{\footnotesize#1}}

\def\mref#1{{\footnotesize ({\ref{#1}})}}

\newtheorem{expl}{Example}[section]

\newtheorem{theorem}[expl]{Theorem}
\newtheorem{corollary}[expl]{Corollary}

\newtheorem{proposition}[expl]{Proposition}

\begingroup
\def\2#1{\ifnum#1<10 0\fi\the#1}
\xdef\isodayandtime{
{\2\day-\2\month-\the\year\space\2{\count0}:%
\2{\count2}}}
\endgroup

\title{\Large\bf Rational Hadamard products via\\ Quantum Diagonal Operators}

\author{{\sc G. H. E. Duchamp, S. Goodenough}   
\rm\thanks{LIPN - UMR 7030
CNRS - Université Paris 13
F-93430 Villetaneuse, France}\\
{\rm and }{\sc K. A. Penson}
\rm\thanks{Laboratoire de Physique Th\'eorique de la Mati\`ere Condens\'ee, 
Universit\'e Pierre et Marie Curie, CNRS UMR 7600
Tour 24 - 2i\`eme \'et., 4 pl. Jussieu, F 75252 Paris Cedex 05, France}           
}
\date{}
\begin{document}

\maketitle
%\isodayandtime

\begin{abstract} We use the remark that, through Bargmann-Fock representation, diagonal operators of the Heisenberg-Weyl algebra are scalars for the Hadamard product to give some properties (like the stability of periodic fonctions) of the Hadamard product by a rational fraction. In particular, we provide through this way explicit formulas for the multiplication table of the Hadamard product in the algebra of rational functions in $\C[[z]]$.\footnote[0]{Partially supported by the ANR CombPhys.}
\end{abstract}

\section{Heisenberg-Weyl algebra and Bargmann-Fock\\ representation}

Let $\H=l^2(\N,\C)$, the standard separable Hilbert space of complex-valued sequences $(\al_n)_{n\in \N}$ such that $\sum_{n\in\N}|\al_n|^2<\infty$ and $e_n=(\delta_{m,n})_{m\in \N}$ be its the canonical Hilbert basis with 
\begin{equation}
\H_0=\bigoplus_{n\in \N} \C e_n	
\end{equation}
the standard dense subspace of $\H$.\\
Using the classical {\it bra} and {\it ket} denotations, one defines the following operators 
\begin{equation}
	a=\sum_{n\geq 1}\sqrt{n}\ |n-1\rangle\langle n|\ ;\ a^\dag=\sum_{n\geq 0}\sqrt{n+1}\ |n+1\rangle\langle n|
\end{equation}
where ``$|n\rangle$'' (resp. ``$\langle n|$'') stand for ``$|e_n\rangle$'' (resp. ``$\langle e_n|$'').
By definition, $a,a^\dag$ are hermitian conjugates and densely defined (because $\H_0$ is a subspace of $dom(a)$ and $dom(a^\dag)$) and one verifies easily that $[a,a^\dag]=Id_{\H_0}$. Due to their proeminent importance in quantum mechanics, there is a huge litterature on these operators and it can be proved that 
\begin{itemize}
	\item the algebra (in $End(\H_0)$) generated by $a,a^\dag$, can be presented as 
$$
HW_\C=\Big< a,a^\dag\ ;\ aa^+-a^+a=1\Big>_{\C-AAU}
$$
where $\C-AAU$ is the category of $\C$-associative algebras with unit. 	
	\item there is no non-zero representation of $HW_\C$ in a Banach algebra
	\item the representation $\be$ on the space of complex formal power series $\C[[z]]$ by operators $\be(a),\be(a^\dag)$ such that $\be(a)(S)=\frac{d}{dz}S\ ;\ \be(a^\dag)(S)=zS$, known as Bargmann-Fock representation,  is faithful
	\item due to the (only) relator $aa^+-a^+a=1$, the family 
	$$\Big((a^\dag)^k a^l\Big)_{k,\, l\geq\, 0}$$
	 is a basis of $HW_\C$ (the expressions w. r. t. this basis are called ``normally ordered'')
	\item denoting $HW^{(e)}_\C$ the subspace generated by the ``monomials'' $\Big((a^\dag)^k a^l\Big)_{k-l=e}$, one has
\begin{equation}\label{gr_alg}
	HW^{(e_1)}_\C.HW^{(e_2)}_\C\subset HW^{(e_1+e_2)}_\C 
\end{equation}
so that the algebra $HW_\C$ is $\Z$-graded (the parameter $e$ is called ``degree'' by algebraists and ``excess/defect'' -according to its sign - by physicists). 
\end{itemize}

Despite of its simple definition $HW_\C$ supports very rich combinatorial studies \cite{BBM,GOF7,GOF8,GOF4,FPSAC07,OPG,GOF6,D20,GOF2,GOF3,GOF5,Varvak}.

\section{Problem of the rational Hadamard table}

The Hadamard product on generating functions was introduced by the French mathematician Jacques Hadamard \cite{Ha} as a shifted convolution product on the one-dimensional torus ($S^1$, the commutative group of angles). This product (denoted $\odot$ in the sequel) amounts to performing the pointwise product on the coefficients of the expanded sequences 
\begin{equation}
	\Big(\sum_{n=0}^\infty a_n z^n\Big)\odot \Big(\sum_{n=0}^\infty b_n z^n\Big):=\Big(\sum_{n=0}^\infty a_nb_n z^n\Big).
\end{equation}
In \cite{Ju}, it is proved that the Hadamard product of an algebraic and a rational series is algebraic (this theorem was extended later to the noncommutative case by Sch\"utzenberger \cite{Sc1} and the latter could be used as a crucial result in \cite{D21}). Hence the algebraic series are a module over the algebra of rational series and it remains the combinatorial problem of giving explicit formulas for the multiplication of two rational series i. e. elements of $K[[z]]$ of the form $\frac{P(z)}{Q(z)}$ with $Q(0)\not=0$. This problem is called here the ``Rational Hadamard product problem''. We can make this problem very precise in the case when the coefficients are taken in an algebraically closed field (for the sake of readability, it will be here taken equal to $\C$, the field of complex numbers) as the algebra of rational series admits 
\begin{equation}\label{rat_lin_basis}
	(z^n)_{n\in \N} ; \Big(\frac{1}{(1-\al z)^m}\Big)_{\al\in \C^*\atop m\in \N^*}
\end{equation}
as linear basis.
  
The aim of this paper is to show how to compute explicitely the multiplication table of this algebra with respect to the former basis using the classical creation/annihilation operators. To this end, we will use the algebra generated by the two operators on $\C[[z]]$ 
\begin{equation}
\be(a^+):	S\ra zS\ ;\ \be(a):	S\ra \frac{d}{dz} S\ .
\end{equation}
These formulas define (on the vector space of complex series) a faithful representation of the Heisenberg-Weyl algebra $HW_\C$.\\
It follows from Eq \mref{gr_alg} that $HW^{(0)}_\C$ is a subalgebra of $HW_\C$, called the algebra of diagonal operators. In order to compute the Hadamard products of the elements of the basis \mref{rat_lin_basis}, one remarks that the diagonal operators are scalars for the Hadamard product. Indeed, as $HW^{(0)}_\C$ is the linear span of the monomials $(a^+)^na^n$, it is sufficient to prove that, for all $S,T\in \C[[z]]$, 
\begin{equation}
\be((a^+)^na^n)[S]\odot T=S\odot \be((a^+)^na^n)[T]	
\end{equation}

which is easily seen by direct computation 
\begin{eqnarray}
\be((a^+)^na^n)[S]\odot T&=& \sum_{m=0}^\infty \scal{\be((a^+)^na^n)[S]}{z^m}\scal{T}{z^m}z^m= 
\sum_{m=0}^\infty \frac{n!}{m!}\scal{S}{z^m}\scal{T}{z^m}z^m\cr
&=&\sum_{m=0}^\infty \scal{S}{z^m}\scal{\be((a^+)^na^n)[T]}{z^m}z^m=S\odot \be((a^+)^na^n)[T]	
\end{eqnarray}

\section{Results}

In this section, we prove that the Hadamard multiplication operator 

\begin{equation}
f\mapsto \frac{1}{(1-z)^{k+1}}\odot f	
\end{equation}
is a diagonal operator and hence, the corresponding multiplication operator by $\frac{1}{(1-\al z)^{k+1}}$ is the composition of a diagonal operator and a dilation (i. e. a substitution $z\mapsto \al z$) which is itself a Hadamard multiplication operator as 
\begin{equation}
	f(\al z)=\frac{1}{1-\al z}\odot f(z)\ .
\end{equation}

\begin{proposition}
For $k\in\N,\ \al\in \C^*$, let $D_{(\al,k)}$ be the operator ``Hadamard multiplication by 
$\frac{1}{(1-\al z)^{k+1}}$''  
\begin{equation}
f\mapsto \frac{1}{(1-\al z)^{k+1}}\odot f	
\end{equation}
Then, one has \\
1) i) $D_{(1,k)}$ is a diagonal operator through the Bargmann-Fock repesentation $\beta$. More precisely 
\begin{equation}
D_{(1,k)}=\be\Big(\sum_{j=0}^k \frac{\binomial{k}{j}}{j!} (a^+)^ja^j\Big)	
\end{equation}
ii) $D_{(\al,0)}$ is the substitution $z\ra \al z$ (automorphism for $\al\not=0$).\\
iii) Due to the associativity of the Hadamard product, one has 
\begin{equation}
D_{(\al,k)}=D_{(\al,0)}\circ D_{(1,k)}=D_{(1,k)}\circ	D_{(\al,0)}	
\end{equation}
\end{proposition}

These formulas on diagonal operators yield following theorem.

\begin{theorem} 
The multiplication table of the algebra of rational power series is the following
\begin{equation}\label{mult_by_monomials}
z^n\odot z^m=\delta_{m,n}z^n\ ;\ z^n\odot \frac{1}{(1-\al z)^{m+1}}=\binomial{-(m+1)}{n}\al^nz^n
\end{equation}
and 
\begin{eqnarray}\label{mult_by_inverses}
\frac{1}{(1-\al z)^{k+1}}\odot \frac{1}{(1-\al z)^{l+1}}=
D_{(\al\be,0)}\Big[\sum_{j=0}^{k} \frac{\binomial{k}{j}}{j!} (l+1)...(l+j)\frac{z^j}{(1-z)^{l+j+1}}\Big]=\cr
D_{(\al\be,0)}\Big[\sum_{j=0}^{k} \frac{\binomial{k}{j}}{j!} (l+1)^{\bar j} \sum_{s=0}^j \binomial{j}{s} \frac{(-1)^{j-s}}{(1-z)^{l+s+1}}\Big]	
\end{eqnarray}
\end{theorem}

\begin{corollary}
The algebra of rational power series $\ncp{\C}{z}^\circ$ is closed under the action of $HW_\C$
\end{corollary}

\section{Concluding remarks}

In a theorem of G. Cauchon (reproved by other means in \cite{D21}, see also \cite{BR}), one has to consider the algebra of rational series in the sense of Neumann-Malcev algebras. In one variable, it is the field of fractions
\begin{equation}
	\frac{P(z)}{Q(z)}\ ;\ P,Q\in \C[z],\ Q\not=0
\end{equation}
within $\C((z,z^1))$, the algebra of Laurent series. This algebra being a subfield of functions $\Z\ra \C$, the Hadamard product (which is in fact the pointwise product) extends at once.\\
It is immediate to see that this field of frations admits the linear basis 
\begin{equation}\label{full_rat_lin_basis}
	(z^n)_{n\in \Z}\ ; \Big(\frac{1}{(1-\al z)^m}\Big)_{\al\in \C^*\atop m\in \N^*}
\end{equation}
and the multiplication table (formulas \mref{mult_by_monomials} and \mref{mult_by_inverses}) have to be extended by zero as regards the products 
$z^n\odot \frac{1}{(1-\al z)^{m+1}}$ with $n<0$. As a consequence, this algebra again is closed under Hadamard products.

\end{document}